\documentclass[10pt,a4paper,oneside]{article}
\usepackage[english]{babel}
\usepackage[dvips]{graphicx}
\usepackage{latexsym}

\def\E{\mathop{\hbox{\rm E}}\nolimits}

\newtheorem{conjecture}{Conjecture}{\bfseries}{\itshape}
\newtheorem{definition}{Definition}{\bfseries}{\rm}
\newtheorem{lemma}{Lemma}{\bfseries}{\itshape}
\newtheorem{theorem}{Theorem}{\bfseries}{\itshape}
\newenvironment{proof}{\noindent\textit{Proof}\ }{\hfill$\Box$}

\begin{document}

\title{Strategy in Ulam's Game and Tree Code\\
Give Error-Resistant Protocols}
\author{Marcin Peczarski\footnote{Supported by the European Community Research
Training Network \textsc{Games} and Polish KBN grant No. 4 T11C 042 25.}\\
\\
Warsaw University, Institute of Informatics\\
ul. Banacha 2, 02-097 Warszawa, Poland\\
e-mail: marpe@mimuw.edu.pl}
\maketitle

\begin{abstract}
\noindent
We present a new approach to construction of protocols which
are proof against communication errors.
The construction is based on a generalization of the well known Ulam's game.
We show equivalence between winning strategies in this game
and robust protocols for multi-party computation.
We do not give any complete theory.
We want rather to describe a new fresh idea.
We use a tree code defined by Schulman.
The tree code is the most important part of the interactive version
of Shannon's Coding Theorem proved by Schulman.
He uses probabilistic argument for the existence of a tree code
without giving any effective construction.
We show another proof yielding a randomized construction which in
contrary to his proof almost surely gives a good code.
Moreover our construction uses much smaller alphabet.
\end{abstract}

\section{Introduction}

We consider a generalization of the communication model introduced
by Yao \cite{Yao}.
Processors $P_1,P_2,\ldots,P_k$ compute a $k$-argument function
$f\colon X_1\times X_2\times\ldots\times X_k\to Y$.
Initially processor $P_i$ knows only the argument $x_i\in X_i$.
In each communication step some of the processors send messages
to some other processors.
Each message consists of bits.
The communication terminates when all the parties know the value
$f(x_1,x_2,\ldots,x_k)$.
The \emph{communication complexity} of the function $f$ is the minimum
number of exchanged bits which are required to solve the problem,
measured on a ``worst'' input.

In this paper we add noise to the communication.
We consider two kinds of noisy binary symmetric channels.
In an \emph{probabilistic channel} bits are incorrectly
transmitted with a given probability $\varepsilon>0$.
In an \emph{adversarial channel} the fraction of
incorrectly transmitted bits is at most $\varepsilon>0$.
We assume that noise can cause bit inversion but bits are
never lost and never any extra bit is added.

Our aim is to construct a protocol which is proof against noise,
computes function $f$ and exchanges $n=\Theta(n_0)$ bits,
where $n_0$ denotes the communication complexity in the case when
the processors are connected by noiseless channels and the constants
implicit in the $\Theta(\cdot)$ notation depend only on $\varepsilon$.
For an adversarial channel the protocol should always return a good result.
For a probabilistic channel it should fail with probability bounded by  
$e^{-\Omega(n_0)}$.
Rajagoplan and Schulman \cite{RajagoplanSchulman, Schulman} show
that this is possible for a probabilistic channel.
Moreover Schulman concludes in \cite{Schulman} about
an adversarial channel in the case of two processors.
But the proofs in those papers are not constructive.

We concentrate on protocols for an adversarial channel.
It should be clear that the robustness against an adversarial channel is
a stronger property than the robustness against a probabilistic channel.
The following Lemma states this precisely.

\begin{lemma}
If a protocol $\cal P$ exchanges $n$ bits and is proof against any number
of errors less than $\varepsilon_0 n$, then the protocol $\cal P$ fails
with probability bounded by $e^{-\Omega(n)}$ when running on a probabilistic
channel with error probability $\varepsilon < \varepsilon_0$.
\end{lemma}

\begin{proof}
Let $X$ be a random variable denoting the number of inverted bits
during a run of $\cal P$ on a probabilistic channel.
Because transmission events are statistically independent
therefore $X$ has binomial distribution with probability $\varepsilon$
of the success in a single event.

Running on a probabilistic channel, $\cal P$ fails with probability
$\Pr(X\geq\varepsilon_0n)$.
From Tchebyshev exponential inequality we have that for any $\lambda\geq0$
$$\Pr(X\geq\varepsilon_0n) \leq
  {\E e^{\lambda X} \over e^{\lambda \varepsilon_0 n}} =
  \left({\varepsilon e^\lambda + 1 - \varepsilon \over
  e^{\lambda \varepsilon_0}}\right)^n.$$

It is sufficient to show that
$f(\lambda) =
 e^{\lambda \varepsilon_0} - \varepsilon e^\lambda - 1 + \varepsilon > 0$
for some $\lambda \geq 0$.
We have $f(0) = 0$ and $f'(0) = \varepsilon_0 - \varepsilon > 0$.
Hence $f(\lambda) > 0$ for some $\lambda > 0$.
\end{proof}

Our construction is based on a generalization of Ulam's game~\cite{Ulam}.
We show the equivalence between a winning strategy
in this game and a robust protocol.
Ulam's game is widely considered in connection with one-direction
data transmission over noisy channel and error correcting codes.
However, its application requires the presence of noiseless
feedback \cite{Balakirski, LawlerSarkissian}, which is not realistic.
We overcome the problem of noiseless feedback
using a tree code introduced by Schulman~\cite{Schulman}.

We hope that the ideas presented in this paper can bring the results of
\cite{RajagoplanSchulman, Schulman} closer to the edge of practical
applicability.
Our approach seems to be more practical because we use much shorter
alphabet and we do not need back steps in protocol
(like in \cite{RajagoplanSchulman, Schulman}), which make it much simpler.

\section{Tree Code}

Let us remind definition of Schulman's tree code from~\cite{Schulman}.

\begin{definition}
\emph{A $d$-ary tree code over alphabet $S$, of distance parameter
$\alpha$ and depth $n$}, is a $d$-ary tree of depth $n$ in which every
arc of the tree is labeled with a character from the alphabet $S$
subject to the following condition.
Let $u$ and $v$ be any two nodes at some common depth $h$ in the tree.
Let $h-l$ be the depth of their least common ancestor.
Let $a_1a_2\ldots a_h$ and $b_1b_2\ldots b_h$ be the concatenation
of the letters on the arcs leading from the root of the tree to $u$ and $v$,
respectively.
Then $H(a_{h-l+1}a_{h-l+2}\ldots a_h,b_{h-l+1}b_{h-l+2}\ldots b_h)\geq\alpha l$,
where the Hamming distance $H$ counts the number of positions $i$ in which
$a_i\neq b_i$.
\end{definition}

\begin{figure}[ht]
\begin{center}
\includegraphics[scale=0.8]{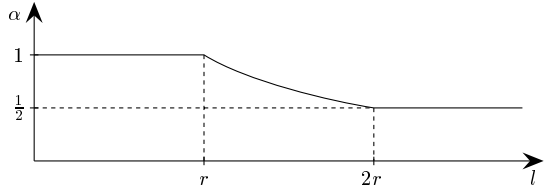}
\end{center}
\caption{The Hamming distance lower bound for the $r$-bit binary tree code.}
\label{hdist}
\end{figure}

We focus on the binary case $d=2$ with distance parameter $\alpha={1\over2}$.
Schulman \cite{Schulman} proves that an alphabet of size 95 suffices to
construct such a code.
The following theorem shows that only 16 letters are sufficient and achieved
Hamming distance is a little better than $1\over2$,
as shown on Figure \ref{hdist}.
Moreover our proof contains a randomized construction which in contrary to the
proof from \cite{Schulman} almost surely gives a good code.
A code over the alphabet $S_r=\{0,1,2,\ldots,2^r-1\}$ is called an $r$-bit code.
We can interpret $S_r$ as the set of all possible $r$-bit vectors.

\begin{theorem}
For every $n\geq1$ and every $r\geq4$ there exists an $r$-bit binary tree code
of depth $n$ satisfying
$$H(a_{h-l+1}a_{h-l+2}\ldots a_h,b_{h-l+1}b_{h-l+2}\ldots b_h)\geq\cases{
l&if $l\leq r$,\cr
r&if $r<l\leq2r$,\cr
l/2&if $l>2r$.\cr
}$$
\end{theorem}

\begin{proof}
Let $t_0,t_1,t_{n-1}$ be a sequence of letters from $S_r$ such that 
for $i<r$ they are linearly independent bit-vectors, e.g. $t_i=2^{r-i}-1$,
and the remaining elements are chosen randomly with uniform distribution.
We label arcs of the tree with alphabet $S_1=\{0,1\}$
so that arcs outgoing from each node to its sons have different labels.
If $a'_1a'_2\ldots a'_h$ is the concatenation of the labels from the root
to a node $u$ over alphabet $S_1$ then as the label over alphabet $S_r$
for the arc leading to the node $u$ from its parent we take
$$a_h=\bigoplus_{i=1}^h a'_i t_{h-i},$$
where $\oplus$ denotes bitwise addition modulo two (exclusive or).

Let $u$ and $v$ be two different nodes at the same depth
and $w$ their least common ancestor.
Let $a_1a_2\ldots a_l$ and $b_1b_2\ldots b_l$ be the concatenations of the
letters over alphabet $S_r$ on the arcs leading from $w$ to $u$ and $v$
respectively.
The Hamming distance between labels leading from the root to $u$ and $v$
is the number of nonzero elements in the sequence $c_1,c_2,\ldots,c_l$,
where $c_i=a_i\oplus b_i$.
If $a'_1a'_2\ldots a'_l$ and $b'_1b'_2\ldots b'_l$ are the concatenations
of the letters over alphabet $S_1$ on the arcs leading from $w$ to $u$
and $v$ respectively and $c'_i=a'_i\oplus b'_i$ then we have
\begin{equation}\label{triangle}
\begin{array}{ccccccccc}
c'_1t_0 &&&&&&&=& c_1\\
c'_2t_0 & \oplus & c'_1t_1 &&&&&=& c_2\\
\vdots &&&&&&&& \vdots\\
c'_lt_0 & \oplus & c'_{l-1}t_1 &\oplus&\ldots&\oplus & c'_1t_{l-1} &=& c_l\\
\end{array}
\end{equation}
Observe that $c'_1=1$.

We can choose $2^{r(n-r)}$ different sequences $t_r,t_{r+1},\ldots,t_{n-1}$.
We denote by $F(n)$ the number of those among them which violate the code
condition (Hamming distance).
Let $f(n,l)$ be the number of sequences which violate the code condition
first time at distance $l$ from the least common ancestor.
Obviously we have
$$F(n)=\sum_{l=1}^n f(n,l).$$

Because $t_0,t_1,\ldots,t_{r-1}$ are linearly independent bit-vectors
and $c'_1=1$ then $c_i\neq0$ for $i\leq r$.
Hence $f(n,l)=0$ for $l\leq2r$.
Therefore thesis holds when $n\leq2r$ or $l\leq2r$.
Subsequently we assume that $n>2r$ and $l>2r$.

If $l$ is even and the code condition is violated for some nodes at distance
$l$ then the number of zeros in the sequence $c_1,c_2,\ldots,c_l$ is
at least $l/2+1$.
Hence the number of zeros in the sequence $c_1,c_2,\ldots,c_{l-1}$ is
at least $l/2$ and the code condition is violated already at distance $l-1$.
Therefore $f(n,l)=0$ when $l$ is even.
Putting $l=2m+2r+1$ we get
$$F(n)=\sum_{m=0}^{\lceil n/2\rceil-r-1}f(n,2m+2r+1).$$

Now we count the number of bad sequences $c_{r+1},c_{r+2},\ldots,c_l$
which can cause code violation first time at distance $l$.
We use a notion of walks on the $xy$-plane, see Figure \ref{walks}.
We assign a walk to each sequence $c_1,c_2,c_3,\ldots$.
The value on the $x$-axis is distance $l$ from the least common ancestor.
The value on the $y$-axis is difference between doubled Hamming distance and $l$.
We start at the point $(0,0)$.
Then we go by the vector $(1,1)$ when succeeding $c_i$ is nonzero and
by the vector $(1,-1)$ otherwise.
The walk corresponds to a bad sequence if it reaches the point $(l-1,0)$ never
previously touching the line $y=-1$ and then it goes to the point $(l,-1)$.
Because $c_i\neq0$ for $i\leq r$ then after the first $r$ moves we are at
the point $(r,r)$.
So the number of bad walks is the number of walks from $(r,r)$ to $(l-1,0)$
such that they never touch the line $y=-1$.
From the reflexion law this is the number of walks from $(r,r)$ to $(l-1,0)$
minus the number of walks from $(r,-r-2)$ to $(l-1,0)$.
The number of walks from $(x_1,y_1)$ to $(x_2,y_2)$ is
$$x_2-x_1\choose(x_2-x_1+y_2-y_1)/2$$
if $x_2-x_1+y_2-y_1$ is even and 0 otherwise.
Consequently there are
$$\displaylines{
{2m+r\choose m}-{2m+r\choose m+r+1}={2m+r\choose m}-{2m+r\choose m-1}=\cr
={2m+r\choose m}-{m\over m+r+1}{2m+r\choose m}={r+1\over m+r+1}{2m+r\choose m}\cr
}$$
bad walks.
In each such walk, after the first $r$ moves up, we go $m$ times up and
$m+r+1$ times down.
The moves up can be chosen on $(2^r-1)^m$ ways because we have $2^r-1$
nonzero choices for each $c_i$.
The move down requires choosing 0 as $c_i$.
Hence we have
$$B(m,r)={r+1\over m+r+1}{2m+r\choose m}(2^r-1)^m$$
bad sequences $c_{r+1},c_{r+2},\ldots,c_l$.

\begin{figure}[ht]
\begin{center}
\includegraphics[scale=0.8]{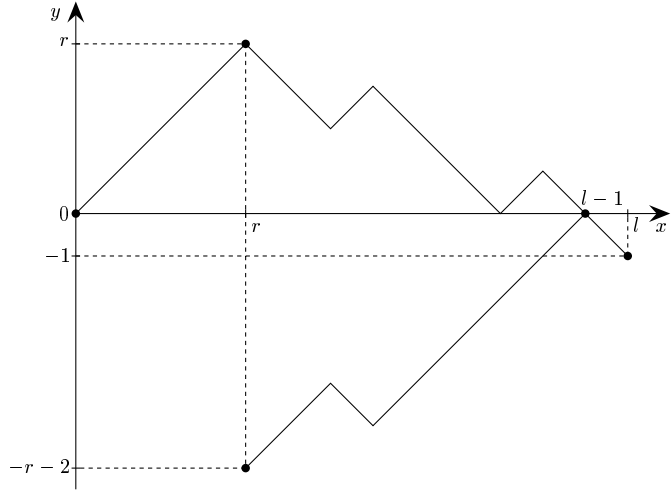}
\end{center}
\caption{A walk on the $xy$-plane.}
\label{walks}
\end{figure}

For every $r\geq1$ there exists $\varepsilon>0$ such that for any $m\geq0$ holds
$${r+1\over m+r+1}\leq{r\over2m+r}2^{1+\varepsilon}.$$
It is sufficient to put $2^\varepsilon={r+1\over r}$.
Therefore we have
$$B(m,r)\leq{r\over2m+r}{2m+r\choose m}2^{rm+1+\varepsilon}.$$

From equations~(\ref{triangle}) we conclude that for every bad choice of
$c_{r+1},c_{r+2},\ldots,c_l$ and for every choice of $c'_1,c'_2,\ldots,c'_l$
there is at most one bad sequence of letters $t_r,t_{r+1},\ldots,t_{l-1}$
which violates the code condition first time at distance $l$.
We have $2^{l-1}$ choices of $c'_1,c'_2,\ldots,c'_l$ because $c'_1=1$.
There are $2^{r(n-l)}$ sequences of $t_l,t_{l+1},\ldots,t_{n-1}$ to choose from.
Therefore
$$\displaylines{
f(n,2m+2r+1)\leq2^{l-1+r(n-l)}B(m,r)=2^{r(n-r)-r^2+r+2(1-r)m}B(m,r)\leq\cr
\leq{r\over2m+r}{2m+r\choose m}2^{r(n-r)-r^2+r+1+\varepsilon+(2-r)m}\cr
}$$
and
$$F(n)\leq2^{r(n-r)-r^2+r+1+\varepsilon}
\sum_{m=0}^{\lceil n/2\rceil-r-1}{r\over2m+r}{2m+r\choose m}2^{(2-r)m}.$$

From~\cite[formulas (5.68), (5.70)]{GKP} we have that
$$\sum_{m=0}^\infty{r\over2m+r}{2m+r\choose m}z^m=
\left(1-\sqrt{1-4z}\over2z\right)^r.$$
We put $z=2^{2-r}$.
For $r\geq4$ we have that $0<z\leq{1\over4}$, $1-\sqrt{1-4z}\leq2z(1+4z)$
and $(1+4z)^r=(1+2^{4-r})^r\leq16$.

Replacing the finite sum with an infinite one we get
$$F(n)\leq2^{r(n-r)-r^2+r+5+\varepsilon}.$$
Finally probability $p$, that randomly chosen sequence
$t_r,t_{r+1},\ldots,t_{n-1}$ violates the code condition, does not exceed
$2^{-r(n-r)}F(n)$.
Hence
$$p\leq2^{-r^2+r+5+\varepsilon}=\hbox{$r+1\over r$}2^{-r^2+r+5},$$
which for $r\geq4$ is less than $0.01$.
\end{proof}

The above proof implies that in practice we can construct a good tree code
with probability as close to $1$ as we need.
There are two possibilities.
Either we choose appropriate $r$
(already for $r=8$ we have $p\leq9\cdot2^{-54}<5\cdot10^{-16}$)
or we use computer to find the first $q$ letters
$t_0,t_1,\ldots,t_{q-1}$ which maximize the Hamming distance.
Then probability $p$ decreases exponentially with~$q$.

In next sections we use the 4-bit binary tree code.
We use the following property of such code.
Assume we are successively receiving letters $a_1a_2a_3\ldots$
describing a path from the root of the tree, and after each letter we
are trying to identify a node to which this path leads.
Then, if the total number of erroneously received letters is at most $r$,
the Hamming distance property of the tree code causes that we badly
identify the node at most $4r$ times.

It should be mentioned that the factor $4$ is rough.
In fact the code described in above proof can have better property.

\section{Generalized Ulam's Game}

Let us remind the definition of Ulam's game.
There are two players Carole and Paul.
The names come from Spencer~\cite{Spencer}.
Carole thinks of a number $x$ between 1 and $n$.
Paul asks at most $q$ questions of the form ``does $x\in F?$'',
where $F\subset\{1,2,\ldots,n\}$.
Carole answers ``yes'' or ``no'' and she is permitted to lie at most $l$ times.
Paul wins if at the end of the game he knows $x$.

We propose a generalization of Ulam's game.
Let $f\colon X_1\times X_2\times\ldots\times X_k\to Y$ be a given function.
Carole thinks of a tuple
$(x_1,x_2,\ldots,x_k)\in X_1\times X_2\times\ldots\times X_k$.
Paul asks at most $q$ questions.
In each question he chooses $i\in\{1,2,\ldots,k\}$ and he asks
``does $x_i\in F_i?$'', where $F_i\subset X_i$.
Like in the original game Carole answers ``yes'' or ``no'' and she
is permitted to lie at most $l$ times.
Paul wins if at the end of the game there is a unique possible value
of the function $f$.

We allow Carole to play an adversary strategy.  Carole does not
actually pick a tuple but answers all questions so that there is
always at least one tuple that satisfies all answers, except for at
most $l$ of them.  We denote the generalized game by ${\cal
G}\left<f,q,l\right>$.  This game is determined.  For given $f,q,l$
either Paul or Carole has a winning strategy.

A winning strategy for Paul is represented by a tree.

\begin{definition}
The \emph{strategy tree} is a labeled binary tree.
Each internal node is labeled with an index $i$ and a set $F_i$.
Arcs leading from each internal node to its sons are labeled with $0$ or $1$.
Leafs are labeled with values of the function $f$.
\end{definition}

The interpretation is as follows.
Paul starts at the root.
In each internal node he asks the question ``does $x_i\in F_i$?''
found in the label of that node.
If the answer is ``yes'' then he goes along the arc with label 1 and
if the answer is ``no'' then he goes along the arc with label 0.
If he reaches a leaf then he reads the value of $f$.
The depth of the strategy tree is the pessimistic number of
questions which have to be asked to win.

As we see in section \ref{ProtocolToStrategy} we can construct a winning
strategy for Paul in the game ${\cal G}\left<f,n_0,0\right>$.
However usually it should be possible to construct a winning strategy
in the game ${\cal G}\left<f,\lceil n_0/(k-1)\rceil,0\right>$ because
intuitively answering one question conforms to sending $k-1$ bits.

For standard Ulam's game, if Carole is required to make sure that
the fraction of lies is less than $1/3$ of her answers
then Paul has a winning strategy with $\Theta(q)$ questions,
where $q=\lceil\log_2n\rceil$ is the number of questions which allows Paul
to win in the game without lies~\cite{SpencerWinkler}.
We expect that for the generalized game a similar result holds.

\begin{conjecture}\label{GameConjecture}
There exists a constant $\varepsilon_g>0$ such that for every function
$f$ for which Paul has a winning strategy in the game
${\cal G}\left<f,n,0\right>$, if the number of lies $l$ is less than the
fraction $\varepsilon_g$ of all questions $q$ $(l<\varepsilon_gq)$,
then Paul has a winning strategy in the game ${\cal G}\left<f,q,l\right>$.
Moreover $q\leq C_g n$, where $C_g$ is a constant which depends only on $l/q$.
\end{conjecture}

Theorem 4 from~\cite{Schulman} and considerations in
section~\ref{ProtocolToStrategy} imply that above conjecture holds
for two-argument functions and with $\varepsilon_g=1/240$.
However we cannot construct a strategy because of a non-constructive
proof in~\cite{Schulman}.
To make a practical benefit from above conjecture a constructive proof
is needed.

Our main result can be stated not so formally as the following
theorem.

\begin{theorem}\label{MainTheorem}
There exists a lie-proof winning strategy for Paul in generalized
Ulam's game with the function $f$
if and only if
there exists an error-proof protocol for $k$ processors computing the
function $f$.
\end{theorem}

Next subsections contain the appropriate constructions which explain the details.

\subsection{Converting Protocol into Winning Strategy}\label{ProtocolToStrategy}

Assume we have a protocol for $k$ processors computing function
$f\colon X_1\times X_2\times\ldots\times X_k\to Y$.
If some transmissions occur at the same time then we order them first
by the number of the sender and next by the number of the receiver.
So we have a protocol in which at every time-step exactly one
processor $P_i$ sends one bit to the another processor $P_j$.
This bit depends on the input $x_i\in X_i$ and on bits which were
seen by $P_i$ before.

Paul simulates the run of the protocol.
Assume that he knows all bits received by all processors up to the present.
So Paul knows the processor $P_i$ which should send a bit at present step.
Hence he can compute the subset $F_i$ of the set $X_i$ such that $x_i\in F_i$
if and only if $P_i$ sends 1 at the present step.
Paul asks ``does $x_i\in F_i$?''.
Answer ``yes'' means that processor $P_j$ receives 1 and
answer ``no'' means that processor $P_j$ receives 0.
If Carole tells the truth then this corresponds to a proper transmission
of the bit.
If Carole lies then this corresponds to the inversion of the bit during
transmission.
Hence Paul knows bit received at present step.
By induction on the step number Paul knows all bits received by all processors
in the run of the protocol.
This sequence of bits is called the \emph{protocol trace}.

Because at the end of the protocol all processors know the same value
of the function $f$ then the protocol trace determines this value and
Paul knows it, as well.
Therefore if the protocol exchanges $q$ bits and is proof against
$l$ errors then we have winning strategy for Paul in the game
${\cal G}\left<f,q,l\right>$.

\subsection{Converting Winning Strategy into Protocol}

Assume we have a winning strategy for Paul in the game
${\cal G}\left<f,q,l\right>$.
We call it shortly a strategy. 
The state of the game is the location in the strategy tree.
We identify the state with the tree node.
If $v$ is a state then by $va$ we denote the son of $v$
to which leads the arc labeled with $a$, where $a\in\{0,1\}$.
Let for a node $v$ the symbols $i(v)$ and $F(v)$ stand for the 
index and set used by Paul in his question,  respectively.
I.e., in node $v$ Paul asks ``does $x_{i(v)}\in F(v)$?''.
Let $[\varphi]$ be 1 if the condition $\varphi$ is true and 0 otherwise.

The strategy corresponds to the following protocol.
All processors maintain the same state.
The initial state is the root.
In the state $v$ processor $P_{i(v)}$ broadcasts bit
$a=[x_{i(v)}\in F(v)]$ to all other processors.
The transmitted bit can be inverted,
but all the processors receive the same value $\hat{a}$.
Moreover we have to assume that there is a feedback to the sender and it
knows what bit has been received by others.
Next all the processors change their state according to the received
bit, to the new state $v\hat{a}$.
If processors reach a leaf then the protocol ends.
It is clear that if the number of faulty broadcasts
does not exceed the number of lies for which the strategy was
developed then all the processors properly compute the value of the
function.

The above scenario is not realistic and does not correspond
precisely to our communication model.
We show how to convert this protocol into another protocol
in which we need neither broadcast nor error-free feedback.

The first problem we have to overcome is the possibility of protocol deadlock.
This happens if $i(v0)\neq i(v1)$ and processors
$P_{i(v0)}$ and $P_{i(v1)}$ receive 1 and 0 respectively.
In this situation all processors will wait for a message forever.
We consider \emph{semi-static strategy} in which all indices at same depth
in the strategy tree are the same.
Every strategy of depth $q$ can be easily converted into a semi-static one,
of depth $q'\leq kq$.
Moreover we can assume that all leafs in the new strategy tree are
at the same depth.
In other cases we can easily extend the strategy tree so that
it has this property.

We label the strategy tree with a tree code using the alphabet $S_4$.
Let $S(v)$ denote the letter on the arc leading to $v$ from its parent.

Now we have the following protocol.
Processor $P_j$ maintains the state $v_j$.
Processor $P_1$ maintains its own state.
Other processors maintain expected state of processor $P_1$.
At beginning all states are the root.

Protocol consists of rounds.
In each round all states are at the same depth in the strategy tree.
Because we consider semi-static strategy then in each round all processors
know the same value $i=i(v_j)$.
In each round processor $P_i$ sends $a=[x_i\in F(v_i)]$ to $P_1$.
Say that $P_1$ receives $\hat{a}$.
If $i=1$ then $P_1$ does not need
to send anything to itself and we assume that $\hat{a}=a$.
Next processor $P_1$ changes its state to $v_1\hat{a}$ and sends
$s=S(v_1\hat{a})$ to all other processors.
Transmission can be corrupted and processors can receive different bits.
Say that for $j\neq1$ processor $P_j$ receives $\hat{s}_j$.
Next $P_j$ updates the state $v_j$ using the tree code property.
Protocol ends when processors reach a leaf.

Let $q_1$ and $q_2$ be the number of messages received and sent by processor
$P_1$, respectively.
Let $r_1$ and $r_2$ be the number of erroneous messages received and sent
by processor $P_1$, respectively.
The protocol works fine if $r_1+4r_2\leq l$.
Hence it is proof against at least $l/4$ bit-errors.
The protocol exchanges $q_1+4q_2$ bits.
We have $q_1\leq q'\leq kq$ and $q_2=(k-1)q'\leq k(k-1)q$.
Hence the protocol exchanges at most $k(4k-3)q$ bits.
Therefore it is proof against at least the fraction
\begin{equation}\label{ach1}
l\over{4k(4k-3)q}
\end{equation}
of badly transmitted bits.

We can increase robustness of the above protocol by running
it parallel on all processors.
Now each processor $P_j$ maintains $k$ states $v_{j1},v_{j2},\ldots,v_{jk}$.
The state $v_{jj}$ is own state of the processor $P_j$.
For $m\neq j$ the state $v_{jm}$ is the state of the processor $P_m$
predicated by the processor $P_j$.

Protocol consists of rounds.
In each round all states are at the same depth in the strategy tree.
Because we consider a semi-static strategy
then in each round all processors
knows the same value $i=i(v_{jm})$.
In each round for each $j\neq i$ the processor $P_i$
sends $a_j=[x_i\in F(v_{ij})]$ to the processor $P_j$.
Say that $P_j$ receives $\hat{a}_j$.
If $j=i$ then $P_i$ does not need to send anything to itself and we assume
that $\hat{a}_i=a_i$.
Next for each $j$ processor $P_j$ changes state to $v_{jj}\hat{a}_j$ and sends
$s_j=S(v_{jj}\hat{a}_j)$ to all other processors.
Say that for $j\neq m$ processor $P_j$ decodes letter from processor $P_m$
as $\hat{s}_{jm}$.
Then $P_j$ updates the state $v_{jm}$ using the tree code property.
Protocol ends when processors reach a leaf.
Each processor has $k$ values of the function.
It is sufficient that at least $\lfloor k/2\rfloor+1$ of them are correct.

\section{Conclusions}

Presented calculations are very rough because our principal aim is to show
that we can construct a robust protocol from a strategy in the game
and we want to show this as simple as possible.
In fact a single bit error by transmitting a letter from the $4$-bit tree code
may not cause an error by decoding this letter.  
Probably in practice a protocol constructed from a good strategy have better
properties than expressed by (\ref{ach1}).

Further work should concentrate on proving Conjecture \ref{GameConjecture}.
Constructive proof which already gives semi-static strategy is desired.
Such a proof would give another proof of 
Schulman's Coding Theorems \cite{Schulman}.
We hope that this approach should give less communication overhead.

Another direction of research is searching for an efficient deterministic
construction of a tree code
and development of an efficient algorithm for tree code decoding.
A ``efficient'' means working in time linear in the depth of the node
in the tree.
Some advice how linear time decoding could be done is given in \cite{Schulman}.

\end{document}